\documentclass[aps,prl,twocolumn,showpacs,amsmath,amssymb,superscriptaddress]{revtex4-1}

\usepackage{graphicx, amsmath, bm, braket, txfonts, color}

\begin{document}

\title{Circuit QED with hole-spin qubits in Ge/Si nanowire quantum dots}

\author{Christoph Kloeffel}
\affiliation{Department of Physics, University of Basel, Klingelbergstrasse 82, CH-4056 Basel, Switzerland}
\author{Mircea Trif}
\affiliation{Department of Physics and Astronomy, University of California, Los Angeles, California 90095, USA}
\author{Peter Stano}
\affiliation{Department of Physics, University of Basel, Klingelbergstrasse 82, CH-4056 Basel, Switzerland}
\affiliation{Institute of Physics, Slovak Academy of Sciences, 845 11 Bratislava, Slovakia}
\affiliation{CEMS, RIKEN, Wako, Saitama 351-0198, Japan}
\author{Daniel Loss}
\affiliation{Department of Physics, University of Basel, Klingelbergstrasse 82, CH-4056 Basel, Switzerland}
\affiliation{CEMS, RIKEN, Wako, Saitama 351-0198, Japan}

\date{\today}

\begin{abstract}
We propose a setup for universal and electrically controlled quantum information processing with hole spins in Ge/Si core/shell nanowire quantum dots (NW QDs). Single-qubit gates can be driven through electric-dipole-induced spin resonance, with spin-flip times shorter than \mbox{100 ps}. Long-distance qubit-qubit coupling can be mediated by the cavity electric field of a superconducting transmission line resonator, where we show that operation times below \mbox{20 ns} seem feasible for the entangling $\sqrt{i\mbox{SWAP}}$ gate. The absence of Dresselhaus spin-orbit interaction (SOI) and the presence of an unusually strong Rashba-type SOI enable precise control over the transverse qubit coupling via an externally applied, perpendicular electric field. The latter serves as an on-off switch for quantum gates and also provides control over the $g$ factor, so single- and two-qubit gates can be operated independently. Remarkably, we find that idle qubits are insensitive to charge noise and phonons, and we discuss strategies for enhancing noise-limited gate fidelities. 
\end{abstract}


\maketitle

In the past decade, the idea of processing quantum information with spins in quantum dots (QDs) \cite{loss:pra98} was followed by remarkable progress \cite{kloeffel:annurev13}. While the workhorse systems are highly advanced, such as self-assembled (In)GaAs QDs \cite{warburton:nat00, gerardot:nat08, godden:apl10, mueller:scirep13, vamivakas:nat10, berezovsky:sci08, press:nphot10, kim:nphys11} and negatively charged, lateral GaAs QDs \cite{petta:sci05, koppens:nat06, nowack:sci07, brunner:prl11, nowack:sci11, gaudreau:nphys12, shulman:sci12}, an emerging theme is the search for systems that allow further optimization. In particular, Ge and Si have attracted attention because they can be grown nuclear-spin-free, which eliminates a major source of decoherence \cite{khaetskii:prl02, merkulov:prb02, coish:prb04}. Promising examples based on Ge/Si are core/shell nanowires (NWs) \cite{lauhon:nat02, lu:pnas05, xiang:nat06, xiang:nna06, hu:nna07, roddaro:prl08, varahramyan:apl09, hao:nlt10, yan:nat11, nah:nlt12, hu:nna12}, self-assembled QDs \cite{wang:procieee07, katsaros:prl11, ares:prl13}, lateral QDs \cite{simmons:prl11, maune:nat12}, and ultrathin, triangular NWs \cite{zhang:prl12}. In addition, spin qubits formed in the valence band (VB, holes) were found to feature long lifetimes \cite{bulaev:prl05, heiss:prb07, gerardot:nat08, fischer:prb08, brunner:sci09, hu:nna12, maier:prb13}. Finally, new sample geometries such as NW QDs are investigated, and have allowed for electric-dipole-induced spin resonance (EDSR) \cite{golovach:prb06} in InAs \cite{nadjperge:nat10, schroer:prl11, petersson:nat12} and InSb \cite{vandenberg:prl13} with spin-flip times down to several nanoseconds only.    

Prime examples for novel qubits are hole spins in Ge/Si NW QDs \cite{hu:nna07, roddaro:prl08, hu:nna12, kloeffel:prb11, maier:prb13}, because they combine all the advantages of group-IV materials, VB states, and strong confinement along two axes. The Si shell provides a large VB offset $\sim$0.5\mbox{ eV} \cite{lu:pnas05}, induces strain, and removes dangling bonds from the core. Furthermore, the holes feature an unusually strong Rashba-type spin-orbit interaction (SOI), referred to as direct Rashba SOI (DRSOI), that is not suppressed by the band gap  \cite{kloeffel:prb11}. We show here that these properties are highly useful for implementing spin qubits. 

In this work, we propose a setup for quantum information processing with holes in Ge/Si core/shell NW QDs. In stark contrast to previous systems \cite{nowack:sci07, golovach:prb06, nadjperge:nat10, schroer:prl11, petersson:nat12, vandenberg:prl13, trif:prb08}, where the EDSR relies on conventional Dresselhaus and Rashba SOI \cite{winkler:book}, the dynamics in our setup are governed by the DRSOI whose origin fundamentally differs. We find that EDSR allows flipping of hole spins within less than \mbox{100 ps}. Two-qubit gates can be realized via circuit quantum electrodynamics (CQED), i.e., with an on-chip cavity \cite{blais:pra04, wallraff:nat04, frunzio:ieeetas05}, where we estimate that operation times below \mbox{20 ns} are feasible for $\sqrt{i\mbox{SWAP}}$. The long-range spin-spin interactions \cite{imamoglu:prl99, trif:prb08, trifunovic:prx12, trifunovic:prx13} enable upscaling. Compared to the original proposal for electron spins in InAs \cite{trif:prb08}, which was recently followed by encouraging results \cite{petersson:nat12}, we find several new and striking features. First, because of bulk inversion symmetry, the SOI and the quantum gates can be precisely controlled by perpendicular electric fields. In particular, these fields serve as on-off switches for two-qubit operations performed on any two spins in the cavity. Second, a strong electric-field-dependence of the $g$ factor allows fine tuning and independent control of all quantum gates. Third, the large DRSOI leads to remarkably short operation times. Finally, we find that noise becomes an issue during quantum operations only, and we discuss how noise-limited gate fidelities can be enhanced. While this paper summarizes our main results, the supplementary information \cite{supplement} (Refs.\ \cite{kloeffel:prb11, winkler:book, lawaetz:prb71, lu:pnas05, luttinger:pr55, luttinger:pr56, lauhon:nat02, xiang:nat06, xiang:nna06, hu:nna07, roddaro:prl08, varahramyan:apl09, hao:nlt10, yan:nat11, nah:nlt12, hu:nna12, blais:pra04, wallraff:nat04, frunzio:ieeetas05, petersson:nat12, mourik:sci12, churchill:prb13, trif:prb08, imamoglu:prl99, blais:pra07, barenco:pra95, loss:pra98, slichter:book, golovach:prl04, borhani:prb06, clerk:rmp10, johnson:pr28, nyquist:pr28, fasth:prl07, nadjperge:nat10, schroer:prl11, meunier:prl07, maier:prb13} cited therein) explains all the derivations and provides the details of the theory.

\begin{figure}[tb]
\begin{center}
\includegraphics[width=0.80\linewidth]{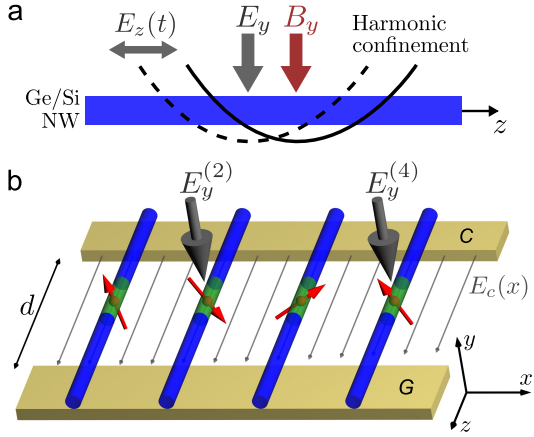}
\caption{Proposed setup. (a) An electric (magnetic) field $E_y$ ($B_y$) is applied perpendicular to the Ge/Si NW QD. Ac fields $E_z(t)$ shift the confining potential along the NW axis $z$. (b) When placed between the center conductor (C) and the ground plane (G) of a transmission line resonator, the hole-spin qubits (red arrows) can interact via the cavity field $E_c$, with the interaction strengths controlled by local electric fields $E_y^{(i)}$. In the sketch, a two-qubit gate is performed between qubits 2 and 4. The setup does not require equidistant QDs and is robust against misalignment.}
\label{fig:Setup3D}
\end{center}
\end{figure} 

Figure \ref{fig:Setup3D}a depicts the setup we consider. Electric gates (not shown) form a Ge/Si NW QD with harmonic confining potential $V(z) = \hbar \omega_g z^2/\bigl(2 l_g^2 \bigr)$ along the wire axis $z$, where $\hbar \omega_g$ is the level spacing, $l_g = \sqrt{\hbar/(m_g \omega_g)}$ is the confinement length, and $m_g$ is the hole mass along $z$ in the subband of lowest energy. An electric (magnetic) field $E_y$ ($B_y$) along $y$ controls the DRSOI (Zeeman splitting). The electric field $E_z$ is induced either externally, $E_z = E_{z,0}^e \cos(\omega_{\rm ac} t)$, or by the cavity, $E_z = E_{z,0}^c (a^\dagger + a)$, where $\omega_{\rm ac}$ is the angular frequency, $E_{z,0}^{e,c}$ is the amplitude, and $a^\dagger$ ($a$) is the creation (annihilation) operator for the quasi-resonant cavity mode \cite{blais:pra04, supplement}. 

When the Ge/Si NW QD of core (shell) radius $R$ ($R_s$) is elongated, $l_g \gg R$, the low-energetic hole states are well described by the Hamiltonian 
\begin{equation}
H = H_{\rm 1D} + V(z) - e E_z z .
\label{eq:Hfull}
\end{equation} 
Here $e$ is the elementary positive charge and $H_{\rm 1D}$ is the effective one-dimensional (1D) Hamiltonian derived in Ref.\ \cite{kloeffel:prb11}. For our setup, one finds $H_{\rm 1D} = H_{\rm LK} + H_{\rm BP} + H_{B} + H_{\rm DR} + H_{\rm R}$, with
\begin{eqnarray}
H_{\rm LK} + H_{\rm BP} &=& A_{+} + A_{-} \tau_z + C k_z \tau_y \sigma_z , \label{eq:LKplusBP} \\
H_B &=& \mu_B B_y \left(- X_2 \sigma_y - X_1 \tau_z \sigma_y + L k_z \tau_x \sigma_z  \right) ,  \label{eq:BalongY} \\
H_{\rm DR} &=& - e E_y U \tau_y .  \label{eq:DRSOI}  
\end{eqnarray}
The Pauli operators $\tau_i$ and $\sigma_i$ act on the transverse band index $\{g,e\}$ and the spin index $\{+,-\}$, respectively. Equation (\ref{eq:LKplusBP}), where $A_\pm = \hbar^2 k_z^2 (m_g^{-1} \pm m_e^{-1})/4 \pm \Delta/2$ and $\Delta = 0.73\mbox{ }\hbar^2/(m R^2) + \Delta_{\rm BP}(\gamma)$, combines the Luttinger-Kohn (LK) \cite{luttinger:pr55, luttinger:pr56} with the Bir-Pikus (BP) Hamiltonian \cite{birpikus:book}, which describe, respectively, the kinetic energy and the effects of strain. The strain-induced energy $\Delta_{\rm BP}(\gamma)$ increases with $\gamma = (R_s - R)/R$, and we note that $\mbox{10 meV} \lesssim \Delta \lesssim \mbox{25 meV}$ in typical Ge/Si NWs with $R \sim \mbox{5--10 nm}$ and $\gamma \sim \mbox{10\%--50\%}$. Equation (\ref{eq:BalongY}), $H_B$, accounts for the orbital effects and the Zeeman coupling due to $B_y$. The SOI comprises the DRSOI $H_{\rm DR}$ induced by $E_y$, Eq.\ (\ref{eq:DRSOI}), and the much smaller standard Rashba SOI (RSOI) $H_{\rm R}$ due to $E_y$ and $E_z$. For the explicit form of $H_{\rm R}$, see Ref.\ \cite{supplement}. The parameters for Ge are \cite{kloeffel:prb11} $C = 7.26\mbox{ }\hbar^2/(m R)$, $U = 0.15 R$, $X_1 \equiv (K+M)/2$, $X_2 \equiv (K-M)/2$, $L = 8.04 R$, $K = 2.89$, $M = 2.56$, $m_g = 0.043 m$, $m_e = 0.054 m$, $m$ is the bare electron mass, and $\hbar k_z = - i \hbar \partial_z$ is the canonical momentum along $z$. 

Our main result is the derivation of the effective 2$\times$2 Hamiltonian for the hole-spin qubit, 
\begin{equation}
H_q = \frac{E_Z}{2} \widetilde{\sigma}_z + T_q \widetilde{\sigma}_x .
\end{equation}
$H_q$ describes the lowest-energy subspace of $H$, Eq.\ (\ref{eq:Hfull}). Its parameters are the Zeeman splitting $E_Z = | g \mu_B B_y | \equiv \hbar \omega_Z$, with $g$ factor $g$, and the transverse coupling $T_q = \bar{\nu} E_z$. Introducing $\nu_{e,c} = \bar{\nu}E^{e,c}_{z,0}$, one obtains $T_q = \nu_e \cos(\omega_{\rm ac} t)$ for EDSR and $T_q = \nu_c (a^\dagger + a)$ for the cavity field. The tilde over the $\widetilde{\sigma}_i$ denotes that the Pauli operators act on the two QD states forming the qubit. Both $E_Z$ and $\bar{\nu}$ are chosen here as positive. The derivation of $H_q$ comprises several basis transformations, two of which we expand perturbatively \cite{supplement}. While the resulting formulas (``model'') for $E_Z$ and $\bar{\nu}$ are too lengthy to be displayed here, they can be very well approximated for realistic Ge/Si NW QDs. Performing a linear expansion in $B_y$ and neglecting $H_{\rm R}$ completely, we find (``approximation'')
\begin{gather}
\bar{\nu} \simeq  \frac{2 E_Z | E_y | e^2 U C}{(\hbar \widetilde{\omega}_g)^2 \widetilde{\Delta}} , \label{eq:nu} \\
g \simeq  2 \left( \widetilde{K} - \frac{L C \Delta}{\widetilde{l}_g^2 \widetilde{\Delta}^2} \right) \exp\Biggl[- \left(\frac{2 e U C E_y}{\widetilde{l}_g \hbar \widetilde{\omega}_g \widetilde{\Delta}} \right)^2\Biggr] , \label{eq:gfactor}
\end{gather}
where
\begin{eqnarray}
\widetilde{K} &=& K - \frac{(K + M) E_y^2}{\left(\frac{\widetilde{\Delta} + \Delta}{2 e U}\right)^2 + E_y^2} \label{eq:Ktilde} ,
\end{eqnarray}
$\widetilde{\Delta} = \sqrt{\Delta^2 + (2 e U E_y)^2}$ is the effective subband spacing,
\begin{equation}
\hbar \widetilde{\omega}_g = \hbar \omega_g \sqrt{ 1 - \frac{2 m_g C^2 \Delta^2}{\hbar^2 \widetilde{\Delta}^3} }
\end{equation}
is the effective level splitting, and $\widetilde{l}_g = l_g \sqrt{\widetilde{\omega}_g / \omega_g}$. Comparing with the exact diagonalization of $H$ (``numerics'') \cite{supplement}, we find that Eqs.\ (\ref{eq:nu}) and (\ref{eq:gfactor}) provide a quantitatively reliable description of the qubit. Considering the complex character of holes and the nontrivial setup with three control fields, the derived formulas are surprisingly simple and therefore provide insight into the role of various parameters. Next, we demonstrate the usefulness of our results by quantifying the basic characteristics of these qubits, such as operation times and lifetimes, and by identifying the most suitable operation schemes.

\begin{figure}[tb]
\begin{center}
\includegraphics[width=0.85\linewidth]{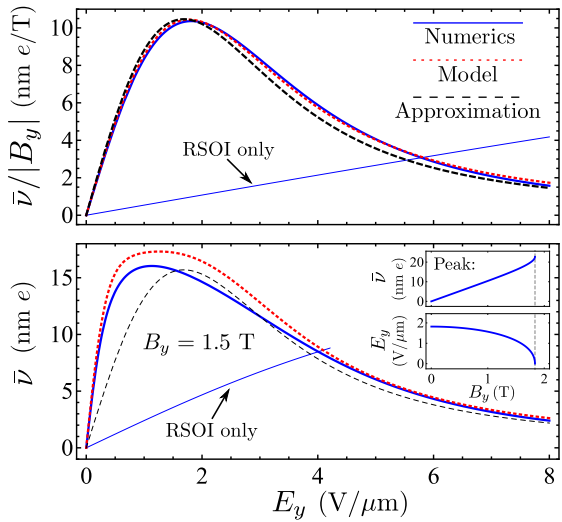}
\caption{Electrical tunability of $\bar{\nu}$ for the QD in the text. Solid blue (dotted red) curves result from the numerical calculation (effective model) \cite{supplement}; dashed black lines correspond to Eq.\ (\ref{eq:nu}). The thin blue lines (RSOI only, $H_{\rm DR} = 0$) illustrate that much stronger $E_y$ would be required for realizing a given $\bar{\nu}$ with the conventional RSOI. Top: Result for $|B_y| \lesssim \mbox{0.5 T}$, where a linear expansion in $B_y$ applies. Bottom: $B_y = \mbox{1.5 T}$, beyond the linear regime. Inset: Height and position of the peak as a function of $B_y$. For $E_y = 0$, a level crossing in the numerics occurs at $B_y \simeq 1.8\mbox{ T}$ (vertical dashed line). The achievable operation times scale with $1/\bar{\nu}$.} 
\label{fig:nu}
\end{center}
\end{figure}

We consider a Ge/Si NW QD with $R = \mbox{7.5 nm}$, $l_g = \mbox{50 nm}$, and $\Delta = \mbox{16 meV}$ based on $R_s \simeq \mbox{10 nm}$. At $E_y = 0$, $g \sim 2K$ and $\hbar \widetilde{\omega}_g = \mbox{0.56 meV} \equiv \hbar \widetilde{\omega}_0$. When $2 K \mu_B |B_y| \ll \hbar \widetilde{\omega}_0$, a linear expansion in $B_y$ applies and both $E_Z \propto |B_y|$ and $\bar{\nu} \propto |B_y|$. In Fig.\ \ref{fig:nu} (top), we plot $\bar{\nu}/|B_y|$ as a function of $E_y$ and find excellent agreement between numerical and perturbative results. The electrical tunability is remarkable. The coefficient $\bar{\nu}$ goes from the exact zero at $E_y = 0$ through a peak at $E_y \simeq \mbox{1.8 V/$\mu$m}$ into a decreasing tail. Most striking is the magnitude, $\bar{\nu}/|B_y| \simeq 10 \mbox{ nm $e$/T}$, which allows for ultrafast single-qubit gates through EDSR. When $\omega_{\rm ac} = \omega_Z$, a $\pi$ rotation on the Bloch sphere requires the spin-flip time $t_{\rm flip} = \hbar \pi / \nu_e$ \cite{kloeffel:annurev13}. For $E_{z,0}^{e} = 10^3\mbox{--}10^4\mbox{ V/m}$ and $B_y = 0.5\mbox{ T}$, for instance, $\nu_e \simeq 5\mbox{--}50\mbox{ $\mu$eV}$ and $t_{\rm flip} \sim \mbox{400--40 ps}$, an extremely short operation time.      

\begin{figure}[tb]
\begin{center}
\includegraphics[width=0.854\linewidth]{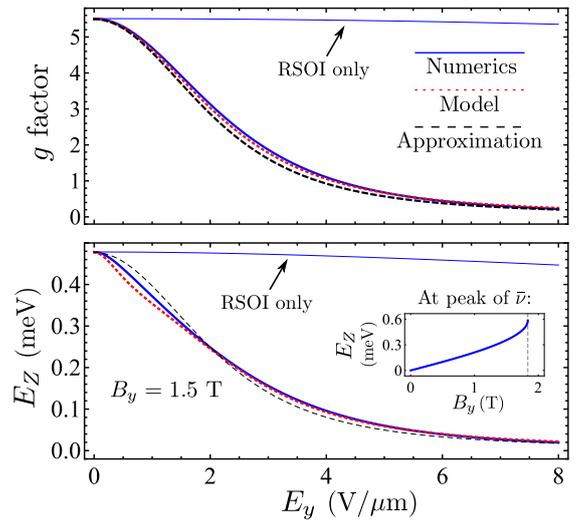}
\caption{The $g$ factor [$|B_y| \lesssim \mbox{0.5 T}$ (top)] and Zeeman splitting [$B_y = \mbox{1.5 T}$ (bottom)] as a function of $E_y$ for the parameters in the text. Solid blue (dotted red) lines are calculated numerically (perturbatively) \cite{supplement}; dashed black lines result from Eq.\ (\ref{eq:gfactor}). The thin blue lines (RSOI only, $H_{\rm DR} = 0$) confirm that the strong electrical tunability results from the DRSOI, Eq.\ (\ref{eq:DRSOI}). Inset: $E_Z$ at the $E_y$ for which $\bar{\nu}$ is maximal (see inset of Fig.\ \ref{fig:nu}).}
\label{fig:Zeeman}
\end{center}
\end{figure} 

The decrease of $\bar{\nu}$ at large $|E_y|$ is related to the $g$ factor decay. As shown in Fig.\ \ref{fig:Zeeman} (top), $g$ decreases from  $g \simeq 5.5$ at $E_y = 0$ toward $g \sim 0$ when $E_y$ is increased to several \mbox{V/$\mu$m}. This tunability was already observed numerically in Ref.\ \cite{maier:prb13}, and our model provides an explanation for the rapid decay of $g$ in this setup. First, as seen in Eqs.\ (\ref{eq:gfactor}) and (\ref{eq:Ktilde}), the main contribution $\widetilde{K}$ changes from $K$ toward a much smaller value $(K-M)/2$ when $E_y$ strongly mixes the subbands $g$ and $e$. In addition, the $g$ factor averages out to zero when the spin-orbit length becomes much smaller than $\widetilde{l}_g$ \cite{flindt:prl06, trif:prb08, maier:prb13}, leading to the exponential suppression. We note that a similar tunability of $g$ was recently measured \cite{ares:prl13} in SiGe nanocrystals.

For the QD under study, the linear expansion applies well for $|B_y| \lesssim \mbox{0.5 T}$ ($2 K \mu_B |B_y| \lesssim \hbar \widetilde{\omega}_0 / 3$). However, it may also be useful to operate the qubit at rather strong $B_y$. In Fig.\ \ref{fig:nu} (bottom), we plot $\bar{\nu}$ for the example $B_y = \mbox{1.5 T}$. As expected, the perturbative results show deviations from the exact calculation as $E_Z$ approaches the orbital level spacing. Nevertheless, they remain correct qualitatively. Compared to $|B_y| \lesssim \mbox{0.5 T}$, the simulated $\bar{\nu}$ peaks at smaller $|E_y|$ and the maximum value, $\bar{\nu} \simeq 16 \mbox{ nm $e$}$, is even greater than the one extrapolated from Fig.\ \ref{fig:nu} (top). For $E_{z,0}^{e} = 10^3\mbox{--}10^4\mbox{ V/m}$, $t_{\rm flip} \sim \mbox{100--10 ps}$. As plotted in the inset of Fig.\ \ref{fig:nu}, the trends found for $B_y = \mbox{1.5 T}$ are enhanced as $B_y$ approaches the value at which neighboring levels cross, allowing the realization of $\bar{\nu} > 20 \mbox{ nm $e$}$. Figure \ref{fig:Zeeman} (bottom) shows that the perturbative results for $E_Z$ are reliable even at high magnetic fields.          

\begin{figure}[tb]
\begin{center}
\includegraphics[width=0.85\linewidth]{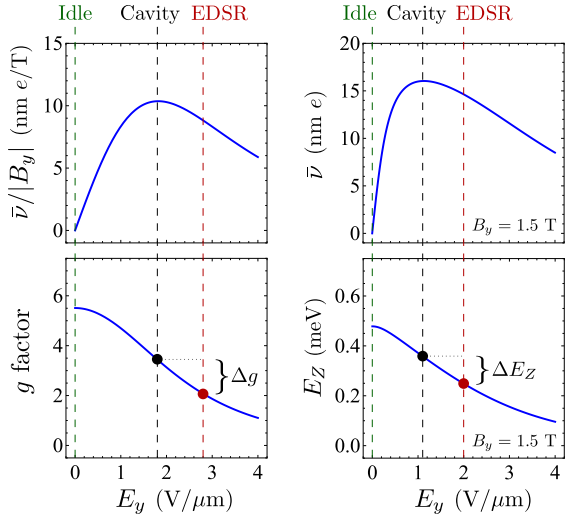}
\caption{Basic operation scheme with the numerical results from Figs.\ \ref{fig:nu} and \ref{fig:Zeeman}. When $E_y = 0$ (idle), the qubit features long lifetimes. Two-qubit operations are envisaged at $E_y$ with maximal $\bar{\nu}$ (cavity). Single-qubit gates can be performed independently by applying a different $E_y$ (EDSR) for which all cavity modes are far off-resonant. The associated change in the $g$ factor ($E_Z$) is indicated by $\Delta g$ ($\Delta E_Z$). For $|B_y| \lesssim \mbox{0.5 T}$ (left), $\bar{\nu}$ is maximal at $E_y \simeq 1.8\mbox{ V/$\mu$m}$, where $g \simeq 3.4$ and $\partial_{E_y} g \simeq -1.6\mbox{ $\mu$m/V}$. For $B_y = \mbox{1.5 T}$ (right), $\bar{\nu}$ peaks at $E_y \simeq 1.1\mbox{ V/$\mu$m}$, where $E_Z \simeq 0.35\mbox{ meV}$ and $\partial_{E_y} E_Z \simeq -0.13\mbox{ nm $e$}$.}
\label{fig:Comparison}
\end{center}
\end{figure}

Thus far, we have identified three major features: a tunable $g$ factor, a strong transverse coupling driven by $E_z$, and precise electrical control via $E_y$. When combined, these features prove ideal for implementing two-qubit gates via CQED. The proposed setup is sketched in Fig.\ \ref{fig:Setup3D}b. Ge/Si NWs are placed perpendicular to the 1D resonator and host a qubit each inside the cavity. Because the $i$th qubit can only couple to the cavity electric field when $E_y^{(i)} \neq 0$, the fields $E_y^{(i)}$ can be used to control qubit-cavity interactions and, hence, two-qubit gates. In addition, the $E_y^{(i)}$ provide precise control over the detunings $\Delta_q^{(i)} = E_Z^{(i)} - \hbar \omega_c$, where $\hbar \omega_c$ is the energy of the cavity mode. This allows the implementation of fast quantum gates through fine tuning of $\Delta_q^{(i)}$. Moreover, as illustrated in Fig.\ \ref{fig:Comparison}, all single- and two-qubit gates can be performed independently.
      
Quantitative information about the cavity field is summarized in Refs.\ \cite{blais:pra04, supplement}. For the mode of lowest energy, we estimate \cite{supplement} that $e E_{z,0}^c/(\hbar \omega_c) \sim 10^{-2}/\mbox{$\mu$m}$ is feasible by decreasing the mode volume compared to Refs.\ \cite{blais:pra04, wallraff:nat04, frunzio:ieeetas05}. From Fig.\ \ref{fig:Comparison}, we deduce $E_Z \simeq 0.35\mbox{ meV}$ at maximal $\bar{\nu}$ for $B_y = 1.5\mbox{ T}$. With $E_{z,0}^c = 3\mbox{ V/m}$, $\nu_c \simeq 50\mbox{ neV}$. Thus, Rabi oscillations in the qubit-cavity system require $\hbar \pi/\nu_c \sim 40 \mbox{ ns}$ for a full cycle at resonance. When $\nu_c^{(i)}/|\Delta_q^{(i)}| < 1$, the coupling between qubits $i$ and $j$ is determined by the transverse spin-spin interaction
\begin{equation}
J_{xy}^{(i,j)} = \nu_c^{(i)} \nu_c^{(j)} \left(\frac{1}{\Delta_q^{(i)}} + \frac{1}{\Delta_q^{(j)}} \right) ,
\end{equation}
which is the basis for the entangling $\sqrt{i\textrm{SWAP}}$ gate \cite{imamoglu:prl99, blais:pra04, blais:pra07, trif:prb08, supplement}. For numerical estimates, we set $\nu_c^{(i)} = \nu_c^{(j)} = \nu_c$, $\Delta_q^{(i)} = \Delta_q^{(j)} = \Delta_q$, and $J_{xy}^{(i,j)} = J_{xy} = 2 \nu_c^2/\Delta_q$. Because corrections to $J_{xy}$ are on the order of $\nu_c^4/\Delta_q^3$ only, we allow for $\nu_c/|\Delta_q|  \simeq \mbox{0.1--0.5}$, which results in short $\sqrt{i\textrm{SWAP}}$ times, $\hbar \pi/|2 J_{xy}| \sim \mbox{100--20 ns}$. At larger $B_y$ (and/or larger $E_{z,0}^c$), these can be reduced further.

In general, qubits that can be manipulated electrically are also sensitive to charge noise \cite{shulman:sci12, kuhlmann:nphys13}. Remarkably, idle qubits in our setup are insulated from the environment; at $E_y = 0 = E_z$, all first derivatives of $E_Z$ and $T_q$ with respect to $E_y$ and $E_z$ vanish, cavity fields are negligible due to off-resonance, and relaxation via phonons is suppressed \cite{maier:prb13} for the magnetic field $\bm{B}$ along $y$. At maximal $\bar{\nu}$, we derive \cite{supplement}
\begin{eqnarray}
1/T_1^{\rm el} &=& 2 \kappa_z^2 \bar{\nu}^2 R_z E_Z /\hbar^2 , \\
1/T_\varphi^{\rm el} &=& \kappa_y^2 \left( \partial_{E_y} E_Z \right)^2 R_y k_B T_y /\hbar^2 , 
\end{eqnarray}
from the Bloch-Redfield theory \cite{slichter:book, golovach:prl04, borhani:prb06} and the spectral functions for Johnson-Nyquist noise \cite{johnson:pr28, nyquist:pr28, clerk:rmp10}. Here $T_1^{\rm el}$ ($T_\varphi^{\rm el}$) is the relaxation (dephasing) time due to electrical noise, $R_\alpha$ ($T_\alpha$) is the resistance (temperature) of the gate that generates $E_\alpha$ along $\alpha \in \{y,z\}$, and the $\kappa_\alpha$ convert fluctuations in the gate voltages to fluctuations in $E_\alpha$. Considering $E_Z > k_B T_\alpha$, we find $T_\varphi^{\rm el} \gg T_1^{\rm el}$ for the values from Fig.\ \ref{fig:Comparison}, which implies $T_2^{\rm el} = 2 T_1^{\rm el} \propto 1/(\kappa_z^2 E_Z^3)$ for the decoherence time. Thus, the setups should be designed such that $\kappa_z$ is small. Assuming $R_\alpha = 10^2\mbox{ $\Omega$}$ and $\kappa_z = \mbox{0.1/$\mu$m}$, we obtain $T_2^{\rm el} = 1\mbox{ ms}$ (30\mbox{ $\mu$s}) for $B_y = 0.5\mbox{ T}$ (1.5\mbox{ T}). If gate fidelities are limited by charge noise, they can be increased by lowering $E_Z$ or $\kappa_z$ or even by operation away from maximal $\bar{\nu}$. If, instead, the fidelities are limited by phonons, they can be much enhanced in the short-wavelength regime at larger $E_Z$ \cite{golovach:prl04, meunier:prl07, trif:prb08, maier:prb13}. Noise that is slow compared to the operation times can be dynamically decoupled \cite{viola:prl99v83, vandersar:nat12, shulman:sci12, kloeffel:annurev13}. 

We studied variants in the setup geometry. For $\bm{B}$ along $x$, $\bar{\nu} = 0$ even at $E_y \neq 0$. Although large $\bar{\nu}$ are possible for $\bm{B}$ along $z$, such a setup requires stronger $\bm{B}$ due to the smaller $g$ factor \cite{maier:prb13, hu:nna12} and exact alignment of all NWs, which is challenging. When the ac fields are perpendicular to the NW, $\bar{\nu}$ becomes several orders of magnitude weaker because of $l_g \gg R$. Hence, the setup we propose in Fig.\ \ref{fig:Setup3D} is the most favorable one.    

\begin{acknowledgments} 
We thank F.\ Maier, K.\ D.\ Petersson, J.\ R.\ Petta, R.\ J.\ Warburton, J.\ R.\ Wootton, and R. Zielke for helpful discussions and acknowledge support from the Swiss NF, NCCRs Nanoscience and QSIT, SiSPIN, DARPA, IARPA (MQCO), S$^3$NANO, SCIEX, the NSF under Grant No.\ DMR-0840965 (M.T.), and QIMABOS-APVV-0808-12 (P.S.).  
\end{acknowledgments}

\end{document}